# Quantum mechanics helps in learning for more intelligent robot

Dao-Yi Dong, Chun-Lin Chen, Zong-Hai Chen, Chen-Bin Zhang

*Department of Automation, University of Science and Technology of China, Hefei, Anhui, 230027, P. R. China*

E-mail: dydong@mail.ustc.edu.cn; clchen@mail.ustc.edu.cn

**Abstract**

A learning algorithm based on state superposition principle is presented. The physical implementation analysis and simulated experiment results show that quantum mechanics can give helps in learning for more intelligent robot.

## 1. Introduction

Quantum computers were proposed in the early 1980's [1] and simultaneously Feynman observed that quantum computers consisting of quantum-mechanical systems have an information-processing capability much greater than that of corresponding classical systems [2, 3]. Several years later Deutsch described a quantum Turing machine and showed that quantum computers could indeed be constructed [4]. Since then there has been extensive research in this field and many results have shown that quantum computers indeed are more powerful than classical computers on some specialized problems, in which Shor's factoring algorithm [5, 6] and Grover's searching algorithm [7, 8] are the most famous two quantum algorithms.

Shor algorithm provides a striking exponential speedup for factoring large integers into prime numbers over the best known classical algorithms and it has also been realized using nuclear magnetic resonance (NMR) [9]. Grover algorithm can achieve a quadratic speedup in unsorted database searching over the best known classical algorithms and its experimental implementations have also been demonstrated using NMR [10, 11, 12] and quantum optics [13, 14]. Recently two directed quantum searching algorithms have been proposed [15, 16] and the results show that the quantum state can monotonically move closer to the target state. Quantum searching algorithm derives its importance from the widespread use of search-based techniques in classical algorithm and has been used to many aspects such as quantum counting, solution of NP-complete problems, search of unstructured databases [17, 18] and quantum control [19].

Here we are inspired by Grover's searching algorithm and propose a learning algorithm based on quantum state superposition principle, which is an important problem in artificial intelligence, especially in robotics. The task of learning is to improve a system's behavior by making it more appropriate for the environment from the point of view of a long-term future performance. This Letter also analyzed the physical implementation and gave a simulated experiment for this quantum learning algorithm (QLA). The theoretical result shows that the needed fundamental transformations to realize our algorithm is the same as that needed in Grover algorithm and is within current state-of-the-art technology. The simulated experiment shows that QLA can help robot learn faster and become more intelligent.

This Letter is organized as follows. Section 2 gives the abstracted problem. In Section 3, we propose a quantum learning algorithm. Section 4 analyzes the physical implementation of this learning algorithm. Section 5 describes a simulated experiment and verifies the effectiveness and superiority of this algorithm. Conclusion is given in Section 6.

## 2. The abstracted problem

In daily life, we always need to learn to make ourselves more intelligent and the aim of artificial intelligence (AI) is to simulate the intelligence of man. In artificial intelligence, designing learning algorithm is an important task. Recently some learning methods including supervised, unsupervised and reinforcement learning (RL) have been proposed. Especially RL has become an important approach to machine learning [20, 21, 22, 23, 24] and is widely used in artificial intelligence [25, 26, 27]. It uses a scalar value named reward to evaluate the input-output pairs and learns by interaction with environment through trial-and-error. In a way, RL is a restatement of the entire AI problem and provides a more generous learning framework for an intelligent robot. In this Letter, we focus on the learning problem of robot based on reinforcement learning.

Now let's consider a typical learning system for a mobile robot. For example, this robot wants to move from Beijing to London while no teacher tells it how to achieve the goal. It must accumulate experience by itself and become more intelligent through learning from its environment. Assume the state of robot is $S$ and related action is $A_{(s_n)}$, which can be divided into discrete values. At a certain step, the robot observes the state of the environment (inside and outside of the robot) $s_t$, and then choose an action $a_t$. After executing the action, the robot receives a feedback $r_{t+1}$, which reflects how good that action is (in a short-term sense). The goal of the learning system is to learn a mapping from states to actions and to make the robot move from Beijing to London with a minimum cost, that is to say, the robot is to learn a policy $\pi : S \times \cup_{i \in S} A_{(i)} \to [0,1]$, so that expected sum of discounted reward of each state will be maximized:

$$\begin{aligned} V^{\pi}_{(s)} &= E\{r_{t+1} + \gamma r_{t+2} + \gamma^2 r_{t+3} + \cdots | s_t = s, \pi\} \\ &= E[r_{t+1} + \gamma V^{\pi}_{(s_{t+1})} | s_t = s, \pi] \\ &= \sum_{a \in A_s} \pi(s,a)[r_s^a + \gamma \sum_{s'} p_{ss'}^a V^{\pi}_{(s')}] \end{aligned} \quad (1)$$

where $\gamma \in [0,1]$ is discounted factor, $\pi(s,a)$ is the probability of selecting action $a$ according to state $s$ under policy $\pi$, $p_{ss'}^{a} = \Pr\{s_{t+1} = s' | s_t = s, a_t = a\}$ is probability for state transition and $r_s^a = E\{r_{t+1} | s_t = s, a_t = a\}$ is expected one-step reward.

In practical applications, there are some difficult problems, such as tradeoff of exploration and exploitation and slow learning speed. Especially when the state-action space becomes huge, the number of parameters to be learned grows exponentially with its size and "the curse of dimensionality" likely occurs [28]. So we explore a quantum learning algorithm and show that it is possible to help a mobile robot faster find an optimal path from Beijing to London using quantum mechanics and overcome some shortcomings of traditional learning methods.

## 3. Learning algorithm

In a quantum learning system, the states may lie in a superposition state:

$$|s^{(m)}\rangle = \sum_{s=00\cdots0}^{\overbrace{11\cdots1}^{m}} C_s |s\rangle \qquad (2)$$

Thus the mapping from states to actions is $f(s) = \pi : S \rightarrow A$, and we have

$$f(s) = |a_s^{(n)}\rangle = \sum_{a=00\cdots0}^{\overbrace{11\cdots1}^{n}} C_a |a\rangle \qquad (3)$$

where $C_s$ and $C_a$ is probability amplitudes of state $|s\rangle$ and action $|a\rangle$, respectively.

Based on the above representation, the QLA can be described as follows.

(1) *Initialize the state and action to the equal superposition state, respectively; i.e.*

$|s^{(m)}\rangle = \frac{1}{\sqrt{2^m}} \sum_{s=00\cdots0}^{\overbrace{11\cdots1}^{m}} |s\rangle$, $f(s) = |a_s^{(n)}\rangle = \frac{1}{\sqrt{2^n}} \sum_{a=00\cdots0}^{\overbrace{11\cdots1}^{n}} |a\rangle$, *and $V(s)$ arbitrarily*

(2) *Repeat the following process (for each episode)*

*For all states* $|s^{(m)}\rangle = \frac{1}{\sqrt{2^m}} \sum_{s=00\cdots0}^{\overbrace{11\cdots1}^{m}} |s\rangle$:

(i). *Observe* $f(s) = |a_s^{(n)}\rangle$ *and get* $|a\rangle$;

(ii). *Take action* $|a\rangle$, *observe next state* $|s'^{(m)}\rangle$, *reward* $r$, *then*

(a) *Update state value:* $V(s) \leftarrow V(s) + \alpha(r + \gamma V(s') - V(s))$

(b) *Update probability amplitudes:*
*repeat Grover operator for* $L$ *times*

$$U_{Grov}^{L} |a_s^{(n)}\rangle = [U_a U_{a_k}]^L |a_s^n\rangle$$

*Until for all states* $|\Delta V(s)| \leq \varepsilon$.

This QLA is inspired by the state superposition principle and the state value can be represented with quantum state. And the occurrence probability of eigenvalue is determined by probability amplitude, which is updated according to rewards. So this approach represents the whole state-action space with the superposition of quantum state and makes a good tradeoff between exploration (trying more inexperienced action to find better policy) and exploitation (taking the most advantage of the experienced knowledge) using probability. As for simulation algorithm on traditional computer it is an effective algorithm with novel representation and computation methods. Simultaneously the representation and computation mode are consistent with quantum parallel computation and can speed up learning in exponential scale with quantum computer or quantum logic gates.

In this learning algorithm we use temporal difference (TD) prediction for the state value updating, and TD algorithm has been proved to converge for absorbing Markov chain when the stepsize is nonnegative and degressive [23]. Since QLA is a stochastic iterative algorithm and Bertsekas and Tsitsiklis have verified the convergence of stochastic iterative algorithms [22], we have the convergence result about the quantum learning algorithm:

**Theorem 1:** *For any Markov chain, quantum learning algorithm converges at the optimal state value function $V(s)^*$ with probability 1 under proper exploration policy when the following conditions hold (where $\alpha_k$ is stepsize and nonnegative):*

$$\lim_{T \to \infty} \sum_{k=1}^{T} \alpha_k = \infty, \qquad \lim_{T \to \infty} \sum_{k=1}^{T} \alpha_k^2 < \infty \qquad (4)$$

## 4. Physical implementation consideration

Now let's consider the physical implementation of the present learning algorithm. In QLA, two key operations are preparing the equally weighted superposition state for initializing the system and carrying out a certain times of Grover iteration for updating probability amplitude according to reward value. These can be implemented using the Walsh-Hadamard transform and the conditional phase shift operation, both of which are relatively easy in quantum computation.

The hadamard transform (or Hadamard gate) is one of the most useful quantum gates and can be represented as [17]:

$$H \equiv \frac{1}{\sqrt{2}} \begin{bmatrix} 1 & 1 \\ 1 & -1 \end{bmatrix} \qquad (5)$$

Through the Hadamard gate, a qubit in the state $|0\rangle$ is transformed into a superposition state of two states, i.e.

$$H|0\rangle = \frac{1}{\sqrt{2}}|0\rangle + \frac{1}{\sqrt{2}}|1\rangle \qquad (6)$$

Similarly, a qubit in the state $|1\rangle$ is transformed into the superposition state $\frac{1}{\sqrt{2}}|0\rangle - \frac{1}{\sqrt{2}}|1\rangle$, i.e. the magnitude of the amplitude in each state is $\frac{1}{\sqrt{2}}$ but the phase of the amplitude in the state $|1\rangle$ is inverted. In classical probabilistic algorithms the phase does not have an analog since the amplitudes are in general complex numbers in quantum mechanics. Now consider a quantum system described by $n$ qubits, it has $2^n$ possible states. To prepare an equally weighted superposition state, initially let each qubit lie in the state $|0\rangle$, then we can perform the transformation $H$ on each qubit independently in sequence and thus change the state of the system. The state transition matrix representing this operation will be of dimension $2^n \times 2^n$ and can be implemented by $n$ shunt-wound Hadamard gates. This process can be represented into:

$$H^{\otimes n}|\overbrace{00\cdots 0}^{n}\rangle = \frac{1}{\sqrt{2^n}} \sum_{a=00\cdots 0}^{\overbrace{11\cdots 1}^{n}} |a\rangle \tag{7}$$

According to the above method, we can accomplish the initialization of state and action. The other fundamental operation required is the conditional phase shift operation which is an important element to carry out the Grover iteration. According to quantum information theory, this transformation may be efficiently implemented on a quantum computer. For example, the transformation describing this for a 2 state system is of the form: $\begin{bmatrix} e^{i\phi_1} & 0 \\ 0 & e^{i\phi_2} \end{bmatrix}$, where $i = \sqrt{-1}$ and $\phi_1$, $\phi_2$ are arbitrary real numbers [7]. The conditional phase shift operation does not change the probability of each state since the square of the absolute value of the amplitude in each state stays the same.

To update probability amplitudes, we reinforce the action corresponding to larger reward value through repeating Grover operator for $L$ times. At first, we initialize the action

$$f(s) = |a_s^{(n)}\rangle = \frac{1}{\sqrt{2^n}} \sum_{a=00\cdots 0}^{\overbrace{11\cdots 1}^{n}} |a\rangle \tag{8}$$

Then we construct a reflection transform

$$U_a = 2|a_s^{(n)}\rangle\langle a_s^{(n)}| - I \tag{9}$$

which preserves $|a_s^{(n)}\rangle$, but flips the sign of any vector orthogonal to $|a_s^{(n)}\rangle$. Geometrically, when $U_a$ acts on an arbitrary vector, it preserves the component along $|a_s^{(n)}\rangle$ and flips the component in the hyperplane orthogonal to $|a_s^{(n)}\rangle$. This also can be looked as an operation of inversion about the mean value of the amplitude [18]. Now change $|a_s^{(n)}\rangle$ with the *k*-th computational basis state $|a_k\rangle$ and construct another reflection transform

$$U_{a_k} = I - 2|a_k\rangle\langle a_k| \qquad (10)$$

Thus we can form a unitary transformation (Grover operator)

$$U_{Grov} = U_a U_{a_k} \qquad (11)$$

From [7,8], we know that by repeatedly applying the transformation $U_{Grov}$ on $|a_s^{(n)}\rangle$, we can enhance the probability amplitude of the *k*-th basis action $|a_k\rangle$ while suppressing the amplitude of all other actions. This can also be looked as a kind of rotation in two-dimensional space [17]. The initial action $f(s)$ can be re-expressed as

$$f(s) = |a_s^{(n)}\rangle = \frac{1}{\sqrt{2^n}}|a_k\rangle + \sqrt{\frac{2^n - 1}{2^n}}|\varphi\rangle \qquad (12)$$

where $|\varphi\rangle = \sqrt{\frac{1}{2^n - 1}} \sum_{a \neq a_k} |a\rangle$. Let us define the angle $\theta$ satisfying $\sin\theta = 1/\sqrt{2^n}$, thus

$$f(s) = |a_s^{(n)}\rangle = \sin\theta |a_k\rangle + \cos\theta |\varphi\rangle \qquad (13)$$

Then from [29], we know that applying the Grover operator $U_{Grov}$ $L$ times on $|a_s^{(n)}\rangle$ can be represented as

$$U_{Grov}^L |a_s^{(n)}\rangle = \sin((2L+1)\theta)|a_k\rangle + \cos((2L+1)\theta)|\varphi\rangle \qquad (14)$$

Through repeating Grover operator, we can reinforce the probability amplitude of corresponding action according to the reward value. Though the operations are similar to that needed in Grover algorithm, their objectives are different since Grover algorithm uses Grover operator to steer an arbitrary state to a specific basis state with a high probability but QLA only applies Grover operator to reinforce some actions. In QLA, the iteration times $L$ is determined according to learning control precise and corresponding reward value.

## 5. Simulated experiment

To evaluate QLA in practice, we simulated a quantum learning system to perform a typical task as the following example describes. A gridworld environment is shown in Fig. 1 each cell of the grid corresponds to an individual state of the environment. From any state the robot can perform one of four primary actions: up, down, left and right, and actions that would lead into a blocked cell are not executed. The task of the learning system is to find an optimal policy which will let the robot move from *S* to *G* with minimized cost (or maximized rewards).

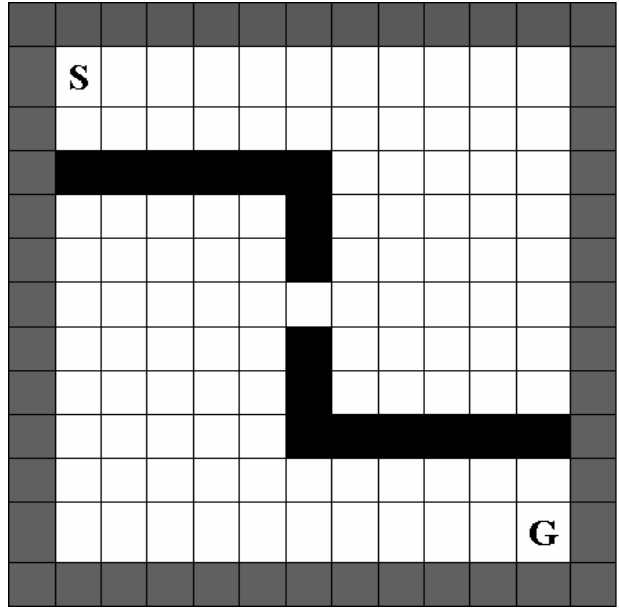

**Fig. 1.** The example of learning task in a gridworld environment with cell-to-cell actions (up, down, left and right). The labels *S* and *G* indicate the initial state and the goal in the simulation experiment described in the text.

In this $13 \times 13 (0 \sim 12)$ grid world, the initial state and the goal is *cell(1,1)* and *cell(11,11)* and before learning the robot has no information about the environment at all. Once the robot finds the goal state it receives a reward of 100 and then ends this episode. All steps are punished by a reward of -1. The discount factor $\gamma$ is set to 0.99 for all the algorithms that we have carried out. In the experiment, we compare QLA with TD(0) and we also demonstrate the expected result on a quantum computer theoretically. For the action selection policy of TD algorithm, we use ϵ-greedy policy (ϵ = 0.01). As for QLA, the action selecting policy, inspired by the collapse theory of quantum measurement, is obviously different from traditional learning algorithms. And the value of $|C_a|^2$ is used to denote the probability of an action $|a\rangle$ defined as $f(s) = |a_s^n\rangle = \sum_{a=00\cdots0}^{11\cdots1} C_a |a\rangle$. For the four cell-to-cell actions $|C_a|^2$ is initialized uniformly.

Just as shown in Fig. 2, a traditional TD algorithm performs about 2000 episodes before finding the optimal policy and the simulated QLA performs better. We observe that QLA is also an effective algorithm on traditional computer although it is inspired by quantum state superposition principle and is designed for quantum computer. QLA explores more than TD at the beginning of learning phrase, but it learns much faster and guarantees a better balancing between exploration and exploitation. In addition, it is much easier to tune the parameters for QLA than for traditional ones. On the contrast, instead of simulating the parallel computing, the theoretically estimated performance of QLA will need only several dozens of episodes to find the optimal policy once practical quantum computer is available. So QLA has great potential of powerful computation in the future, which will lead to a more effective approach for the existing problems of learning.

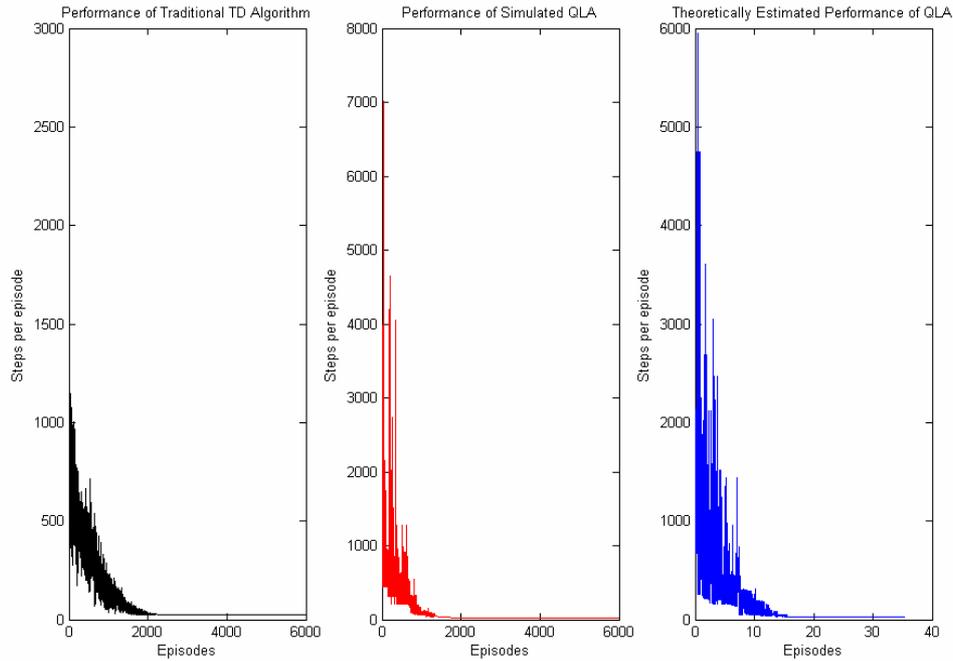

**Fig. 2.** Performance of QLA compared with traditional TD algorithm (ϵ-greedy policy) and the expected theoretical result on quantum computer.

All the results show that QLA is effective and excels traditional learning algorithms in the following two aspects: (1) Action selecting policy makes a good tradeoff between exploration and exploitation using probability, which speeds up the learning and guarantees the searching over the whole state-action space as well. (2) Representation is based on the superposition principle and the updating process is carried through parallel computation, which will be much more prominent in the future when practical quantum computer comes into use.

## 6. Conclusion

In this Letter we are inspired by Grover's searching algorithm and propose a learning algorithm based on quantum state superposition principle. We also analyze its physical implementation and give some simulated experiments. The theoretical result shows that the needed fundamental transformations to realize our algorithm is the same as that needed in Grover algorithm and is within current state-of-the-art technology. The simulated results show that quantum mechanics can give helps in learning for more intelligent robot. Our work also shows that quantum computers indeed are more powerful than classical computers following Shor algorithm and Grover algorithm. Moreover, this result demonstrates that quantum mechanics can speed up machine learning and be used to artificial intelligence.

## Acknowledgements

This work has been supported by the Innovation Research Project of Science and Society Practice of the Chinese Academy of Science for Graduate Students (No. 2004-18), and the Innovation Fund of the University of Science and Technology of China for Graduate Students (KD2004048).